\documentclass[twocolumn, deluxetables]{aastex631}
\usepackage{amsmath}
\usepackage[caption=false]{subfig}
\usepackage{multirow}
\usepackage{graphicx}

\graphicspath{{./}{figures/}}

\shorttitle{XRISM Resolves the Fe~K band in NGC 4151}
\shortauthors{XRISM Collaboration}

\begin{document}

\title{XRISM Spectroscopy of the Fe~K$_{\alpha}$ Emission Line in the Seyfert AGN NGC~4151\\ Reveals the Disk, Broad Line Region, and Torus}


\author{XRISM Collaboration}
\affiliation{Corresponding Author: Jon M. Miller, jonmm@umich.edu}

\author[0000-0003-4721-034X]{Marc Audard}
\affiliation{Department of Astronomy, University of Geneva, Versoix CH-1290, Switzerland} 

\author{Hisamitsu Awaki}
\affiliation{Department of Physics, Ehime University, Ehime 790-8577, Japan} 

\author[0000-0002-1118-8470]{Ralf Ballhausen}
\affiliation{Department of Astronomy, University of Maryland, College Park, MD 20742, USA} 
\affiliation{NASA / Goddard Space Flight Center, Greenbelt, MD 20771, USA}
\affiliation{Center for Research and Exploration in Space Science and Technology, NASA / GSFC (CRESST II), Greenbelt, MD 20771, USA}

\author[0000-0003-0890-4920]{Aya Bamba}
\affiliation{Department of Physics, University of Tokyo, Tokyo 113-0033, Japan} 

\author[0000-0001-9735-4873]{Ehud Behar}
\affiliation{Department of Physics, Technion, Technion City, Haifa 3200003, Israel} 

\author[0000-0003-2704-599X]{Rozenn Boissay-Malaquin}
\affiliation{Center for Space Science and Technology, University of Maryland, Baltimore County (UMBC), Baltimore, MD, 21250 USA}
\affiliation{NASA / Goddard Space Flight Center, Greenbelt, MD 20771, USA}
\affiliation{Center for Research and Exploration in Space Science and Technology, NASA / GSFC (CRESST II), Greenbelt, MD 20771, USA}

\author[0000-0003-2663-1954]{Laura Brenneman}
\affiliation{Center for Astrophysics | Harvard-Smithsonian, MA 02138, USA} 

\author[0000-0001-6338-9445]{Gregory V.\ Brown}
\affiliation{Lawrence Livermore National Laboratory, CA 94550, USA} 

\author[0000-0002-5466-3817]{Lia Corrales}
\affiliation{Department of Astronomy, University of Michigan, MI 48109, USA} 

% ...

\begin{abstract}
We present an analysis of the first two XRISM/Resolve spectra of the well-known Seyfert-1.5 active galactic nucleus in NGC 4151, obtained in December 2023.  Our work focuses on the nature of the narrow Fe~K$_{\alpha}$ emission line at 6.4~keV, the strongest and most common X-ray line observed in AGN.  The total line is found to consist of three components.  Even the narrowest component of the line is resolved with evident Fe~K$_{\alpha,1}$ (6.404~keV) and K$_{\alpha,2}$ (6.391~keV) contributions in a 2:1 flux ratio, fully consistent with neutral gas with negligible bulk velocity.  Subject to the limitations of our models, the narrowest and intermediate-width components are consistent with emission from optically thin gas, suggesting that they arise in a disk atmosphere and/or wind.  Modeling the three line components in terms of Keplerian broadening, they are readily associated with (1) the inner wall of the ``torus,'' (2) the innermost optical ``broad line region'' (or, ``X-ray BLR''), and (3) a region with a radius of $r\simeq 100~GM/c^{2}$ that may signal a warp in the accretion disk.  Viable alternative explanations of the broadest component include a fast wind component and/or scattering; however, we find evidence of variability in the narrow Fe~K$_{\alpha}$ line complex on time scales consistent with small radii.  The best-fit models are statistically superior to simple Voigt functions, but when fit with Voigt profiles the time-averaged lines are consistent with a projected velocity broadening of FWHM$=1600^{+400}_{-200}~{\rm km}~{\rm s}^{-1}$.  Overall, the resolution and sensitivity of XRISM show that the narrow Fe~K line in AGN is an effective probe of all key parts of the accretion flow, as it is currently understood.  We discuss the implications of these findings for our understanding of AGN accretion, future studies with XRISM, and X-ray-based black hole mass measurements.

\end{abstract}


\section{Introduction}
New techniques have succeeded in imaging the inner and outer extremes of accretion flows onto massive black holes.  The Event Horizon Telescope has revealed the shadow of the event horizon in M87 and Sgr A* (\citealt{ehtc19}, \citealt{ehtc22}), while IR and sub-mm observations have imaged molecular gas on scales of 0.1--1~pc \citep{grav20}.  In contrast, most of the geometry within the roughly 5--6 decades of physical scale between these extremes -- and the processes that drive accretion disks and winds over this expanse -- must be probed using spectroscopy, variability, and other techniques (see, e.g., \citealt{chen1989}, \citealt{peterson2004}; also see \citealt{gmc23}).

The proximity and resultant high flux observed from the Seyfert-1.5 galaxy NGC 4151 have placed it at the forefront of all such efforts ($d = 16$~Mpc, \citealt{bk15}; or $z = 0.0033$, \citealt{bentz06}).   NGC 4151 was among the first galaxies wherein delays between continuum flux from the central engine and the flux in broad optical lines located the responding gas \citep{gaskell86}.  This provided a means of measuring black hole masses by translating Keplerian line broadening -- dependent upon radius in units of $r = GM/c^{2}$ -- to a physical distance given by $r \propto c\tau$.  Optical broad line region (BLR) reverberation is now a common means of measuring black hole masses (see, e.g., \citealt{peterson2004}, \citealt{cackett2021}).   A recent, robust reverberation lag measured in NGC 4151 gives a black hole mass of $M_{BH} = 3.4^{+0.4}_{-0.4}\times 10^{7}~M_{\odot}$ \citep{bk15}.

Observations continue to reveal additional details of the BLR, including radial and vertical gradients and structures.  In the context of narrow, neutral Fe~K emission lines, it is particularly interesting to note that {\em low-ionization} optical lines tend to show little or no bulk shift and may originate closer to the disk surface, while higher ionization lines are shifted and may be excited downstream within a wind (\citealt{cs88}, \citealt{kollatschny2003}).  Some broad optical lines -- potentially those observed at low inclinations -- display double-peaked profiles, indicating a strong association with the disk \citep{eracleous2003}.

Of particular interest is the relationship between the BLR and the dusty molecular ``torus'' geometry that divides Type-1 and Type-II AGN in unification schemes (e.g., \citealt{antonucci1993}).  The name of this structure is simply meant to signal an axially symmetric region that subtends a large solid angle.  The geometry that is now clear in protoplanetary disks -- that of an inner gas disk and outer dust disk (e.g., \citealt{henning2013}) -- likely approaches a correct physical picture.  Dust reverberation studies in quasars, for instance, find that the inner wall of the torus is likely only a few times larger than the optical BLR \citep{minezaki2019}.  This supports a picture wherein the BLR and inner torus may be facets of a transition region rather than entirely distinct.  This is also true in NGC 4151, where 1--4~$\mu$m reverberation mapping finds signals from 0.033~pc and 0.076~pc \citep{lyu2021}; the smaller value corresponds to a radius of approximately $r \sim 3\times 10^{4}~GM/c^{2}$.

Narrow Fe~K$_{\alpha}$ emission lines from neutral gas are the clearest atomic lines seen in X-ray spectra of AGN (see, e.g., \citealt{gmc23}).  The ubiquity of these lines and their association with cold gas naturally associates Fe~K$_{\alpha}$ lines with the torus.  Although the torus may extend over tens of parsecs, the rapid fall of flux with radius means that most of the line flux likely originates from the {\em inner wall} of the torus.  Some models for the optical BLR predict that radiation pressure on dust, wherein Fe can be embedded, may be important to launching the optical BLR (see, e.g., \citealt{czerny2015}).  Such models suggest a natural physical link between the BLR and torus.  Chandra grating spectroscopy of Seyferts revealed that the width of some Fe~K$_{\alpha}$ lines is consistent with the optical BLR, but found no clear trend between Fe~K$_{\alpha}$ and optical H$\beta$ line widths \citep{shu2010}, though the X-ray and optical data were generally not contemporaneous (also see \citealt{minezaki2015}). 

In particular, Chandra grating spectroscopy of NGC 4151 revealed asymmetry in the narrow Fe~K$_{\alpha}$ line, consistent with weak special relativistic and gravitational shifts from approximately $r \simeq 1000~GM/c^{2}$ -- slightly smaller than the optical BLR -- and tentative evidence of line variability from a region as small as $r \simeq 100~GM/c^{2}$ \citep{miller2018}.  A study of Fe~K$_{\alpha}$ line variability in NGC 4151 found that the line lags the continuum by $\tau = 3.3^{+1.8}_{-0.7}$~days (about half of the H$\beta$ delay), consistent with the radius inferred in fits to time-averaged Chandra spectra \citep{zoghbi2019}.   More recent work has also found a lag signal in the narrow Fe~K$_{\alpha}$ line in NGC 3516 \citep{noda2023}.  Both AGN are often classified as Seyfert-1.5, potentially signaling a particularly revealing line of sight, but they are also among the brightest AGN in X-rays with the strongest narrow Fe~K$_{\alpha}$ lines.  To the extent that they represent the class as a whole, these results may be early evidence that narrow Fe~K$_{\alpha}$ lines are composite features that sample a broad range of radii within the accretion flow.

Herein, we present an analysis of two high-resolution spectra of NGC 4151 obtained using Resolve \citep{resolve22}, the cryogenically cooled microcalorimeter spectrometer aboard X-ray Imaging and Spectroscopy Mission (XRISM; \citealt{xrism20}).  The spectra achieve a resolution of approximately $\Delta E \simeq 4.5$~eV across the 1.6--17.4~keV pass band, giving $R = E/\Delta E \simeq 1420$ at the Fe~K$_{\alpha}$ line.   This resolution is 10 times sharper than the Chandra/HEG (in velocity units, approximately $\Delta v = 200~{\rm km}~{\rm s}^{-1}$ versus $\Delta v = 2000~{\rm km}~{\rm s}^{-1}$).  In the Fe~K band, the collecting area of Resolve is approximately $A=160~{\rm cm}^{2}$ (with the gate valve closed), whereas the area of the Chandra/HEG is approximately $A=30~{\rm cm}^{2}$.  The figure of merit for line detection is FOM$=\sqrt{R\times A}$, so Resolve is $\simeq8$ times more powerful than the HEG in the Fe~K band per unit exposure time.

Section 2 briefly describes the XRISM observations and reduction of the Resolve calorimeter data.  In Section 3, we detail our spectral fits and results.  Section 4 of this paper discusses the strengths and weaknesses of our models and results, consequences for our understanding of AGN accretion flows and X-ray spectra, and lays out future studies with XRISM. Finally, Section 5 restates our findings as distinct conclusions.

\section{Observations and Data Reduction}
XRISM Observation 000125000 started on 02 December 2023 at 06:22:50 (UTC), and observation 000137000 started on 26 December 2023 at 18:32:51 (UTC); hereafter, they are referred to as 125 and 137. In both observations, NGC 4151 was observed without a filter in the Resolve light path.  The Xtend instrument (\citealt{hayashida2018}) was operated in 1/8 window mode for observation 125, but part of observation 137 was done in full window mode. (The 1/8 window mode limits photon pile-up compared to full window mode; the details of the Xtend data are deferred to a subsequent paper.)  To reduce the data from these observations, we followed the procedures described in the XRISM Quick Start Guide Version 2.1 using pre-release tools and calibration files that will soon be made public. 

The Resolve microcalorimeter consists of a 6x6 array of independent pixels that deliver a resolution of 4.0--5.0~eV.  Throughout an observation, the energy of a Mn~K$_{\alpha}$ line that is excited by an onboard $^{55}$Fe source is tracked, enabling gain corrections and event energy corrections through HEASOFT tools.  For the two observations treated in this paper, the gain uncertainty is less than 0.1~eV.  Additional screening was performed to ensure pulse shape validity and to avoid pulse rise-time and frame-time coincidences.  Time filtering was performed to avoid periods corresponding to SAA passage, to avoid Earth's sunlit limb, and to avoid the 50~mK adiabatic demagnetization refrigerator (ADR) recycling intervals (the recycling periods are 4300 s in duration, followed by 42.5 hours of normal operation; one recycling interval occurred during observation 125, no recycling intervals occurred within observation 137).  After filtering, the net exposure times for observations 125 and 137 were 72.4~ks and 55.8~ks, respectively, yielding total count rates of $2.468\pm  0.006~{\rm counts}~{\rm s}^{-1}$ and $3.643\pm  0.008~{\rm counts}~{\rm s}^{-1}$.  

Calorimeter events have different ``grades'' depending on their resolution; this analysis is confined to the highest-resolution primary events, or ``Hp'' events.  In general, the spectra in this analysis have an energy resolution of 4.5~eV.  Response files were generated using the XRISM RMF and ARF simulator tools, using the fully cleaned event file and pre-release files from the XRISM CalDB version 8 database.  The RMF tool is capable of producing responses with very different sizes, retaining relatively more or less information on line spread functions; we have used ``large'' responses.  This choice retains high fidelity at low energy, effectively separating line flux on the wings of the line spread function from continuum flux.  The event energies were not barycenter-corrected.

A full characterization of the XRISM/Resolve background is in progress.  Unlike CCD cameras with a broad field of view, ``off-source'' backgrounds are extremely difficult to obtain with Resolve.  It is clear that the total background includes lines such as Cr K$_{\alpha}$, Mn K$_{\alpha}$, Ni~K$_{\alpha}$, and potentially Cu~K$_{\alpha}$, and that these lines may be enhanced during solar flare activity through solar-wind charge exchange and/or flux that causes materials in the spacecraft to emit.  However, the flux of NGC 4151 is at least 50 times higher than the estimated background flux across the full pass band and 300 times above background at Fe~K$_{\alpha}$, and the resolution of the calorimeter is able to separate lines at the red-shift of NGC 4151 from local lines with zero shift.  We have therefore proceeded to analyze the Resolve data without explicitly accounting for the largely negligible background.  At the time of writing, the Resolve gate valve remains closed, preventing the passage of X-rays with an energy below 1.6~keV.  The reduction procedure and calibrations utilized in this work are therefore optimized for this scenario.  

\section{Analysis and Results} 

\subsection{Sofware and Models}
All spectral fits were made within SPEX \citep{kaastra1996}, minimizing the Cash statistic \citep{cash1979}.  All errors reported in this work correspond to $1\sigma$ confidence intervals.  To avoid small lingering response uncertainties close to the gate valve transmission threshold, we set a lower fitting bound of 2.4~keV.  Eventually, the Resolve response may be characterized out to 20~keV, but we have utilized the response out to 17.4~keV, giving an initial fitting band of 15~keV.  The best-fit parameter values and errors are listed in Table 1, and are based on fits made after binning the Resolve spectra by a factor of 9.0.  This binning achieved a balance between resolution and signal in individual lines.

We adopted a continuum consisting of a $T=25,000$~K blackbody to model the UV accretion disk, and a cut-off power-law that is artificially bent to zero at low energy using the ``etau'' component in SPEX.  The blackbody is included in the fit to ensure a realistic ionization balance in photoionization modeling of the wind absorption spectrum with ``pion'' (\citealt{miller2015}, \citealt{mehdipour2016}).  The blackbody normalization was determined through separate fits to an archival Hubble/STIS spectrum (the blackbody does not contribute emission within the XRISM band, and the related photoionized absorber fits are detailed in Xiang et al., in prep.).   The total continuum was modified with the ``hot'' ISM absorption model to account for the well-known low-energy column density within NGC 4151 (the Milky Way only contributes only $N_{H} = 2.1\times 10^{20}~{\rm cm}^{-2}$, \citealt{hi4pi}); we left all parameters at their default values apart from the total equivalent hydrogen column density, ${\rm N}_{\rm H}$, and the covering factor of this column, $f_{cov}$.  

Figure 1 shows spectra from Obs 125 and 137 in energy space, and in velocity space relative to the neutral Fe~K$\alpha$ line assuming a weighted centroid energy of 6.40~keV.  Even prior to fitting the lines, it is apparent that there are at least three distinct components in each line profile: (1) a very narrow component wherein the Fe~K$_{\alpha,1}$ and K$_{\alpha,2}$ doublet is visible despite being velocity broadened; (2) an intermediate component that is roughly 2--3 times as broad as the narrow component; and (3) a broader, asymmetric extension to approximately 6.0~keV at the base of the line complex.  It is worth noting that observation 137 is brighter and that the structure within its line profile is more distinct; the profile of observation 125 likely indicates a slightly higher degree of velocity broadening (see Table 1).  Even in the narrow components, the broadening is high enough that the fine details of the intrinsic atomic line shapes are unimportant to the fits (\citealt{yaqoob2024}).  It is possible that a much broader component -- potentially relativistic reflection from the innermost accretion disk -- is present in the data; this analysis is deferred to a forthcoming paper including simultaneous NuSTAR and NICER data (Boissay-Malaquin et al.\ 2024).

Building on fits to deep Chandra grating spectra that revealed asymmetry and broadening consistent with Keplerian orbital motion \citep{miller2018}, we elected to fit the line profile with three ``mytorus'' line components, each broadened by an independent ``Speith'' function within SPEX.  The ``mytorus'' component is ideally suited to Resolve data: it has a native resolution of 2~eV, and it self-consistently includes K$_{\alpha,1}$ and K$_{\alpha,2}$ lines, associated K$_{\beta}$ lines, and a Compton shoulder with a strength that depends on the column density of the emitting gas \citep{murphy2009}.  Formally, the parameters of ``mytorus'' include (1) the column density of the emitting gas, (2) the inclination of the emitting gas with respect to our line of sight, (3) the photon index of the illuminating power-law function, (4) the red-shift of the emitting gas, and (5) a flux normalization (in the same units as the power-law normalization; we used the ``mytorus'' table model file corresponding to a spectrum that is cut-off at 100~keV).  Solar abundances were assumed in all fits with ``mytorus'' functions.

The ``Speith'' function is similar to ``rdblur'' or ``relconv'' within XSPEC (\citealt{fabian1989}, \citealt{dauser2013}, \citealt{arnaud1996}); it is nominally able to handle any level of black hole spin, and complicated emissivity profiles (the SPEX implementation is based on \citealt{speith1995}).  We fixed the black hole spin to $a = 0.7$, loosely based on \citet{reynolds2021}, assumed a Euclidean emissivity of $q=3$ (taking $J(r) \propto r^{-q}$) as we are treating gas far from the black hole, and fit for the inner radius (in units of $GM/c^{2}$) and inclination.  For each component, we linked the inclination of the ``mytorus'' and ``Speith'' functions, and linked the power-law index of the ``mytorus'' function to that of the continuum.  In total, we independently fit for the inner line production radius, inclination, gas column density, and line flux normalization in each component.  

Chandra gratings spectra recorded Fe~XXV and Fe~XXVI absorption lines from ``warm absorber'' winds in NGC 4151 and Seyferts (e.g., \citealt{kraemer2005}, \citealt{couto2016}); the Resolve spectra detect Fe~XXV and Fe~XXVI but also reveal narrow absorption lines from less ionized charge states in the 6.45--6.65~keV region in NGC 4151.  Our total model therefore consists of the absorbed continuum as described above, the three narrow Fe~K line components, and multiple zones of photoionized absorption with the ``pion'' model.  Although Fe~K emission line components and absorbers must be included within the same model in order to achieve a good description of the data (including the continuum flux level), the resolution of the spectra effectively isolates the respective influences of the emission and absorption components.   (We note that the undulations seen in the red wing of the Fe~K$_{\alpha}$ line in observation 125 are likely noise; they are not as evident in observation 137, and would require extreme red-shifts or extreme abundances.)  

\subsection{Detailed Modeling}

With the broad continuum and wind absorption determined in preliminary broad-band fits, we restricted the energy range of our subsequent fits to the 5.4--7.4~keV band.  Table 1 details the results of fits to both observations, and the fits are shown in Figure 1. The errors were derived while allowing all Fe~K emission line component parameters and continuum parameters to vary.  Figure 2 depicts modifications to the best-fit model for observation 137, to illustrate the sensitivity of the data to inner production radius and gas column density.

The innermost extent of the narrowest line component is $r_{nar} = 0.56^{+0.50}_{-0.31}\times 10^{4}~GM/c^{2}$ in observation 125, and $r_{nar} = 2.1^{+4.6}_{-1.0}\times 10^{4}~GM/c^{2}$ in observation 137.  The $1\sigma$ errors only just exclude each other.  Both narrow components require very low inclination angles, $\theta_{nar} = 12.0^{+8.2}_{-4.6}$~degrees and $\theta = 10.2^{+7.5}_{-3.8}$~degrees, and the two are formally consistent.  We measure inner radii of $r_{intm} = 3.1^{+1.4}_{-1.2}\times 10^{3}~GM/c^{2}$ and $r_{intm} = 2.8^{+0.5}_{-0.7}\times 10^{3}~GM/c^{2}$ for the intermediate line components in observations 125 and 137.  Not only are these radii formally consistent, but so too are the derived inclinations: $\theta_{intm} = 33.7^{+27.3}_{-7.7}$~degrees and $\theta_{intm} = 24.8^{+8.2}_{-5.1}$~degrees, respectively.  Finally, fits to the ``broad'' component in each spectrum also return consistent results: $r_{brd} = 73.8^{+103.2}_{-35.3}~GM/c^{2}$ and $r_{brd} = 97.0^{+96.3}_{-9.8}~GM/c^{2}$, and $\theta_{brd} = 19.4^{+1.3}_{-0.2}$~degrees and $\theta_{brd} = 19.5^{+0.8}_{-0.5}$~degrees, for observations 125 and 137.  The consistency of the radii and inclinations in independent observations with significant total flux differences suggests that the fitting results are driven by major physical characteristics of the accretion flow.   

We note that our assumed emissivity index, $q=3$, is appropriate for an isotropic source and flat disk.  If an emissivity index of $q = 2$ is assumed instead, the derived radii and inclinations are within the errors of the $q=3$ measurments, though the inclinations drift to nominally higher values.  If we instead assume a flat emissivity of $q=0$ -- corresponding to an extreme corona that blankets the disk and emits equally at all radii -- the Cash statistic increases by $\Delta C = 80$ and $\Delta C = 50$ for observations 125 and 137, respectively.  Apart from being rejected statistically, these fits require inclinations that differ markedly between the components, to the point of being unphysical (e.g., 15 degrees vs 80 degrees).  Subject to the limitations of our models, the data appear to exclude a flat emissivity and radially extended corona.

\subsection{From Velocity Components to Geometry}

Prior work has noted that the dust sublimation radius in NGC 4151 and other AGN likely sets an outer envelope on narrow Fe~K$_{\alpha}$ line production \citep{gandhi2015}.  Factoring in errors, the Resolve data allow for radii of $3500~GM/c^{2} \leq r_{nar} \leq 67000~GM/c^{2}$.  Assuming a black hole mass of $M_{BH} = 3.4^{+0.4}_{-0.4}\times 10^{7}~M_{\odot}$ \citep{bk15}, this equates to $2\times 10^{16}~{\rm cm} \leq r_{nar} \leq 38\times 10^{16}~{\rm cm}$, or $0.005~pc \leq r_{nar} \leq 0.12~pc$.  These values are formally consistent with 1--4$\mu$m dust reverberation lags that estimated inner radii of 0.033~pc and 0.076~pc for the inner wall of the torus \citep{lyu2021}, and also broadly consistent with optical and IR interferometric constraints \citep{gravity2023}.  {\em The narrow component of the Fe~K$\alpha$ line in NGC 4151 likely traces the inner face of the dusty torus.}  

Recent studies of the the optical and UV BLR in NGC 4151 measure line widths of $FWHM \simeq 7000~{\rm km}~{\rm s}^{-1}$ for C III] and Mg II], and $FHWM = 5000~{\rm km}~{\rm s}^{-1}$ for H$\beta$ \citep{bk15}.  Assuming strictly Keplerian motion and an intermediate inclination of $\theta = 45$~degrees, these values correspond to radii of $r = 5000-10,000~GM/c^{2}$.  The intermediate components detected with Resolve point to radii about two times smaller, consistent with $r = 3000~GM/c^{2}$.  Given the complexity of the structure within the Fe~K region, it is not surprising that the line flux errors are relatively large.  However, the flux normalizations of the intermediate component are well determined, and indicate statistically significant variability between observations 125 and 137, with a fraction of 30--50\%. This level of variability is similar to that observed with XMM-Newton and Suzaku when a preliminary reverberation lag of $\tau = 3.3^{+1.8}_{-0.7}$~days was first measured in the narrow Fe~K$\alpha$ line \citep{zoghbi2019}.  {\em The intermediate component of the Fe~K$\alpha$ line in NGC 4151 likely traces the inner extent of the optical/UV BLR, or potentially an ``X-ray BLR.''} 

When interpreted only in terms of Keplerian motions, the broadest line component within the composite Fe~K$_{\alpha}$ emission line suggests a region with a radius of $r \simeq 100~GM/c^{2}$.  While this could be associated with the inner part of a radially truncated accretion disk, this would be inconsistent with the multi-wavelength variability observed in NGC 4151, and with strong UV continuum flux that is likely produced in a standard thin disk that extends close to the black hole.  It could instead be a warp, giving a local enhancement to the solid angle of the disk.  A warp might be caused by radiation pressure \citep{pringle1996}, or by a misalignment between the angular momentum of the accretion flow and the spin axis of the black hole \citep{nixon2012}.  Such a region may potentially play a key shielding role if two reprocessing sites exist within AGN accretion flows (\citealt{gardner2014}, \citealt{edelson2017}).  Multi-wavelength studies of NGC 4151 find that the O III ionization cone and the axis of the inner jet resolved with the VLBI are misaligned with the large-scale radio jet by 12--14 degrees, potentially also suggesting a warp \citep{evans1993}.

However, the broadest component of the Fe~K$_{\alpha}$ line complex may not be shaped by orbital motion, and at least two alternatives could potentially produce the observed red wing.  First, although Compton scattering by free electrons gives rise to a very sharp Compton shoulder, scattering by bound electrons could give rise to a smoother red wing (\citealt{odaka2016},  \citealt{tanimoto2019}).  Second, the red wing could originate through bulk motion rather than radial motion, if the line is produced in a wind accelerating to velocities far in excess of the local escape velocity (see, e.g., Hagen et al.\ 2024, in prep.).  In both cases, the production region could be more consistent with the BLR.  

\subsection{Line Variability}

Line variability can provide an independent angle on where each component is produced.  The top panels in Figure 3 show the XRISM/Resolve light curve of observations 125 and 137 in the 2--10~keV band.  In rough terms, each light curve divides into low-flux and high-flux halves.  We therefore made spectra from the low and high flux halves of each observation, and created ``difference'' spectra by subtracting the low flux phase from the high flux phase.  Features that vary on time scales comparable to the segments are revealed in a model-independent manner in difference spectra.  The segments are 70~ks and 50~ks in duration, for observations 125 and 137, respectively.  These durations correspond to distances of $d = c\Delta t = 260-470~GM/c^{2}$ in NGC 4151 assuming direct light paths, or distances a factor of a few smaller including likely geometrical considerations (e.g., complex light echoes from a cylindrical geometry; see \citealt{welsh1991}, \citealt{grier2012}).  

The bottom panels in Figure 3 show the difference spectra after rebinning by a factor of 50 within SPEX.  We are only interested in the narrow Fe~K$_{\alpha}$ line, but weaker emission and absorption features are evident in the 6.5--7.0~keV band that indicate the warm absorber winds may also vary on these time scales, potentially placing them within hundreds of gravitational radii from the black hole.  

We fit a simple power-law plus Voigt line plus power-law model to each spectrum (fixing the Gaussian and Lorentzian components to have the same width).   In observation 125, evidence of a response in the Fe~K$\alpha$ line is relatively strong.  The centroid of the variable line flux is E$=6.38\pm 0.01$~keV, and the normalization is $\tau = -2.9\pm 0.5$, indicating a feature significant at the $5\sigma$ level of confidence (within SPEX, a negative optical depth functions as a flux normalization in emission).  In this observation, the variable line flux is resolved and gives a {\it projected} velocity width of $FWHM = 3800^{+1400}_{-900}~{\rm km}~{\rm s}^{-1}$.  When fit using the ``Speith'' model blurring a delta function, the best fit gives $r = 100^{+700}_{-20}~GM/c^{2}$ and an inclination of $\theta = 11\pm 1$~degrees.  The low inclination leads to a low projected velocity from a small radius with high intrinsic velocities; when the inclination is fixed at $\theta = 19$~degrees as per the results in Table 1, the best-fit radius is $r = 350\pm 100~GM/c^{2}$, but the overall fit is worse by $\Delta C = 15$.   In observation 137, we find weak evidence of narrow Fe~K$\alpha$ emission at an energy of E$=6.28^{+0.06}_{-0.02}$~keV.  The normalization is $\tau = -1.8^{-2.4}_{+0.6}$, indicating a feature that is nominally significant at roughly the 3$\sigma$ level of confidence.  These results offer support for an element of the composite line originating at fairly small radii, but scattering at larger radii is not entirely excluded.

\subsection{Comparison with Voigt Profiles}

To obtain a measure of the extent to which our three-component models are statistically preferred over the Voigt profiles that are often fit to optical emission lines from AGN,  we made alternative fits to the Fe~K region using the Voigt line model within SPEX.  To replicate the Fe~K$_{\alpha}$ and K$_{\beta}$ emission lines modeled with ``mytorus,'' two Voigt functions were included for each spectrum, fixed at the appropriate energy values for neutral gas.  The free parameters then included the ``optical depth'' of each emission line, the FWHM of the Gaussian component, and the FWHM of the Lorentzian component.  For consistency, the $K_{\beta}$ Gaussian and Lorentzian widths were initially fixed to the corresponding values for the $K_{\alpha}$ line.   

The results of fits with this model are shown in Figure 4.  The Voigt models are simpler, with $\Delta\nu = 6$ additional degrees of freedom in each fit, but the Voigt profiles fail to capture the complexity of the data and {\it increase} the Cash statistic by $\Delta C = +105$ and $\Delta C = +95$ for observations 125 and 137, respectively, over the fits reported in Table 1.  If the Cash statistic is treated as a $\chi^{2}$ statistic and an F-test is used to calculate a false alarm probability, the best-fit models are improvements over the Voigt models at more then the 8$\sigma$ level of confidence.  The increases in the Cash statistic are only smaller by 10 if the Gaussian and Lorentz widths are decoupled within the Voigt profile.  The best-fit width for both observations is consistent with FWHM$=0.035^{+0.008}_{-0.005}$~keV, or FWHM$=1600^{+400}_{-200}~{\rm km}~{\rm s}^{-1}$.  Although these values do not reflect the richness of the data, they may enable straightforward comparisons to optical and UV lines.

\subsection{Additional Constraints}

Beyond the results listed in Table 1, we note that allowing the red-shift of the narrowest Fe~K emission components to vary did not improve the fit to either spectrum.  The shift of the narrow and intermediate Fe~K$\alpha$ line components is $z \leq 2\times 10^{-4}$.  This confirms four important things: (1) the emitting gas is likely neutral, since ionized charge states of Fe have different centroid energies and different separations between their K$_{\alpha}$ and K$_{\beta}$ lines; (2) the red-shift of the accretion flow is fully consistent with the red-shift of the host galaxy, likely ruling out binarity, and indicating that the black hole is close to the dynamical center of the host galaxy (alternatively, if the black hole orbits the dynamical center of its host, we must be viewing the black hole close to apasteron); (3) the geometries sampled by these components are likely axially symmetric with respect to the black hole; (4) these components likely trace the base of an outflow or an atmosphere, rather than a region downstream, as the gas has not yet been accelerated to a significant velocity shift.

\section{Discussion}
We have analyzed the Resolve calorimeter spectra from the first two XRISM observations of the Seyfert-1.5 AGN NGC 4151.  By virtue of its proximity, NGC 4151 is the brightest radio--quiet AGN in X-rays above E$=$4~keV, with the strongest narrow Fe~K$_{\alpha}$ emission line.  It was included in the XRISM Performance Verification phase with the expectation that spectroscopy of this source would most clearly reveal the accretion flow between the innermost disk and torus in unobscured AGN, and establish what will be possible in longer observations of this source class.  
Fitting the Fe~K$_{\alpha}$ lines with high-resolution models that include the necessary atomic structure, we find that the ``line'' is actually a composition of emission lines that are concretely associated with the inner face of the torus and the innermost extent of the optical BLR (or, ``X-ray BLR''), and tentatively with a warp or structure at a radius of approximately $r \simeq 100~GM/c^{2}$.  The narrowest line components are consistent with neutral Fe, have negligible bulk velocity, and may favor an origin in Compton-thin gas.  Taken together, these findings suggest that the emission lines may be generated at the low-velocity base of winds launched from the leading edge of these geometries.

\subsection{Geometrical Implications}

Although our results are likely the clearest demonstration that some narrow Fe~K$_{\alpha}$ line flux in unobscured AGN originates in the optical BLR, or at slightly smaller radii, they are not the only evidence.  Continuum and Fe~K$_{\alpha}$ line flux variability observed in NGC 4151 and NGC 3516 was found to contain lags consistent with light travel times between the central engine and BLR (\citealt{zoghbi2019}, \citealt{noda2023}).  Spectroscopy of the deepest Chandra observations of NGC 4151 also revealed asymmetries consistent with emission close to the optical BLR, and even variability consistent with a responding geometry at $r = 100~GM/c^{2}$ \citep{miller2018}.  

The revolutionary aspect of Resolve spectroscopy and sensitivity is the ability to clearly {\em separate} velocity components within Fe~K$_{\alpha}$ lines, enabling physical modeling of the conditions in different line production regions.  Our models are necessarily incomplete, but the data have permitted initial measurements of the innermost production radius, inclination, and optical depth of three distinct regions spanning a factor of 100 (or more) in distance from the central engine -- scales that are not accessible with direct imaging in the X-ray, UV, and optical wavelengths that are natural in this regime.

In Seyfert-1.5 AGN, the line of sight to the central engine is likely to just skim the envelope of outflows from the BLR and torus.  This is a particularly appealing explanation of some ``changing look'' behavior -- correlated changes in obscuring column and source flux -- that are observed in NGC 4151 and other sources \citep{bentz22}.  It is also a good explanation of eclipse events that are sometimes observed in Seyfert-1.5 AGN (e.g., \citealt{gallo2021}).  This scenario implies that the torus and BLR both cover a fairly large solid angle, implying significant vertical extent above the plane of the disk.  

Our results sum to a self-consistent view of the accretion flow geometry that supports this picture of Seyfert-1.5 AGN, and unified models of AGN appearance generally (e.g., \citealt{antonucci1993}).  In the broadest sense, the detection of distinctive velocity components {\em requires} enhanced emission from specific radii, consistent with local changes in scale height (covering factor).  A flat or smoothly varying disk profile gives rise to a smooth line profile.  The inclinations that we have measured in each velocity component are consistent between observations 125 and 137.  The fact that all of the inclinations are small reflects the opening angle of the BLR and torus.   If the symmetry axis of the total accretion flow is inclined to our line of sight by $\theta_{los}$ but the BLR or torus have opening angles of $\theta_{open}$, then we would expect to measure $\theta_{meas} = \theta_{los} - \theta_{open}$.  The fact that we measure the lowest inclination from the narrowest line component may signal that the torus has the largest opening angle with respect to the disk, and that the portions of the BLR and potentially warped disk that are traced by Fe~K$_{\alpha}$ lines have smaller opening angles.  This picture is consistent with the ``bowl'' geometry indicated in VLTI/GRAVITY interferomtric data (\citealt{gravity2024}).

Figure 5 shows a cartoon representation of the narrow Fe~K$_{\alpha}$ line production regions within the accretion flow in NGC 4151.  Its depiction of the BLR region is qualitatively similar to the geometry that \citet{bentz22} recently inferred by considering optical lines and variability.  Owing to $F \propto r^{-2}$ flux dilution leading to stronger emission from geometries that are closer to normal with respect the central engine, evidence of emission from predominantly optically thin media, and a potential association with acceleration at the base of the BLR, the line flux is depicted at the face of the geometrical boundaries.  Our results do not suggest that line production is confined to these regions; rather, all regions and radii with cold gas (and/or dust) contribute some line flux.  

\subsection{The Emitting Gas (and Dust)}

Both the narrow and intermediate components of the Fe~K$_{\alpha}$ line are nominally Compton-thin, with $N_{H} < 1.6\times 10^{24}~{\rm cm}^{-2}$.  For this reason, it is appealing to associate the bulk of the line flux from these velocity components with the disk atmosphere or a wind -- potentially the same thing -- rather than an optically thick part of the accretion flow.  Given the negligible bulk velocity observed in these lines, the emission must stem from the base of a flow, before significant acceleration occurs. This is broadly consistent with optical spectra of the BLR, wherein low ionization lines have low velocities and are associated with wind launching regions (\citealt{cs88}, \citealt{kollatschny2003}).  

One family of models suggests that the heightened opacity in the optical and ultraviolet afforded by dust may be essential in lifting gas within the BLR to be accelerated into an outflow (\citealt{czerny2015}).  Neutral Fe~K lines like those in NGC 4151 may trace this process since neutral iron can be embedded within dust.  However, because dust is extremely efficient at blocking optical and ultraviolet radiation, a dusty medium may be optically thick at these wavelengths while still being optically thin in hard X-rays.  In fact, a $10^{23}$ cm$^{-2}$ medium would reach an optical depth of about 100 at 2200 \AA\ (\citealt{weingartner2001}), compared to about 0.5 around 6~keV, with elastic scattering on dust grains being the main contributor to the total opacity at these energies (\citealt{draine2003}).

\subsection{Implications for Other AGN Types}

It may be that Seyfert-1.5 AGN afford an especially fortuitous view of the accretion flow, enabling contributions from different geometries to be disentangled by virtue of an intermediate inclination angle with respect to our line of sight.  This has the effect of making velocity components more distinct than in systems viewed comparatively face-on, and provides a view of the inner edge of the torus.  In contrast, if Compton-thick AGN are viewed close the equatorial plane of the system, the narrowest Fe~K$_{\alpha}$ lines in these systems may trace the extended torus but not the innermost face.  Indeed, in NGC 1068, it is possible that some Fe~K$_{\alpha}$ line flux originates in an extended outflow that may or may not be directly associated with the torus \citep{kallman2014}.  For such reasons, the narrowest Fe~K$_{\alpha}$ velocity components in NGC 4151 and other Seyferts may differ significantly from those that will be observed in highly obscured and/or Compton-thick AGN.

Low-luminosity AGN (LLAGN) and LINER galaxies are powered by accretion onto black holes at a low Eddington fractions.  Ultraviolet flux from the innermost accretion disk is greatly reduced, potentially due to truncation, and in some cases the BLR and/or torus are absent \citep{ho2008}.  The low flux levels that define these AGN may make it difficult to detect multiple velocity components within Fe~K$_{\alpha}$ emission line profiles, but these lines may be more direct probes of the accretion flow than low-flux optical and UV lines that are more readily entangled with emission from the host galaxy.  The most natural test case is M81*, which is also part of the XRISM PV program.

\subsection{Toward Reverberation Mapping}
In comparing the width of Fe~K$_{\alpha}$ lines observed in Seyferts with Chandra to the width of optical H$\beta$ lines, \citet{shu2010} drew attention to the possibility that some line flux may originate in the BLR.  That work did not reveal a correlation between Fe~K$_{\alpha}$ line width and H$\beta$ line width, but most of the data were not taken simultaneously, and even simultaneous data may not account for potential lags between the respective emission regions.  Moreover, the resolution and sensitivity afforded by the Chandra/HETGS necessarily mixed emission from different parts of the accretion flow.  

Resolve clearly makes it possible to isolate the velocity component that is best associated with the BLR.  Even without simultaneous optical spectroscopic monitoring, a comparison of what we have called the ``intermediate'' Fe~K$_{\alpha}$ velocity component in Resolve spectra of a set of unobscured AGN may reveal a correlation with optical H$\beta$ line widths.  A better path forward will be to obtain spectroscopic monitoring in optical bands every time XRISM observes an AGN. An optimal path forward would be to monitor a set of AGNs with both XRISM and optical spectroscopy.  The reward may be a strong correlation that would enable estimates of black hole masses from single-epoch Fe~K$_{\alpha}$ velocity component widths in future integral field unit surveys of AGN obtained with Athena \citep{barret2018} or other missions.

\subsection{Limitations}
A key limitation of our analysis is that we have fit all three components of the ``narrow'' Fe~K$_{\alpha}$ line using a model construction that only considers Keplerian broadening.  Although the bulk radial velocity in the two narrower components appears to be negligible -- potentially suggesting an association with the base of an outflow that has not yet been accelerated -- the broadest component extending down to roughly 6.0~keV could be different.  This line flux could represent emission from a wind that is being accelerated to a high velocity (Hagen et al.\ in prep.).  In this case, its shift would indicate the wind velocity and the flow could be launched from radii much larger than $r\simeq 100~GM/c^{2}$.  

Alternatively, different scattering mechanisms could create this feature. 
Scattering in hot gas that helps to confine neutral Fe that may only be present in cold gas clumps could give rise to weak, broad wings similar to those observed in quasar broad lines \citep{shields1981}.  A third possibility is that the smooth red wing extending down to roughly 6.0~keV is the result of Compton scattering via electrons that are bound to atoms or molecules, rather than free electrons (see, e.g., \citealt{odaka2016}, \citealt{tanimoto2019}).  The tentative detection of variability in the flux of the Fe~K$_{\alpha}$ line on the scales consistent with just $few \times 10^{2}~GM/c^{2}$ weakly favors a geometrical interpretation over wind acceleration or scattering (see Figure 3).  Line flux variations consistent with this radius were also detected with Chandra when NGC 4151 was in a similar high flux state \citep{miller2018}.  Additional observations and analysis are necessary to fully test all of these possibilities.

Finally, the analysis we have presented is based on an early characterization of the instrument, and the first set of models that are suited to such data.  The calibration of the Resolve line spread function and backgrounds will continue to progress, and including these effects will likely deliver small improvements over our measurements.  Future refinements in models may also be able to improve upon our results; we look forward to an extremely high-resolution reflection model that includes emission from all abundant elements, and predicts emission from a broad range of ionization parameters.  

\subsection{Future work}
In the near future, a subset of the XRISM science team will submit papers based on several other aspects of the observations of NGC 4151.  These will include (1) a study of X-ray warm absorbers, disk winds, and evidence of ultra-fast outflows (Xiang et al., in prep.), (2) a comparison of the narrow Fe~K emission line complex with contemporaneous optical line profiles (Noda et al., in prep.), (3) a search for relativistic reflection from the innermost disk, utilizing simultaneous NuSTAR data (Boissay-Malaquin et al.,, in prep), (4) a study of metal abundances in the gas that feeds the black hole in NGC 4151, utilizing simultaneous Chandra spectra (Buhariwalla et al., in prep.), and (5) broadband reflection modeling, utilizing simultaneous NuSTAR and INTEGRAL data (Paltani et al., in prep.).  

Over the course of the PV phase, XRISM will make a total of five observations of NGC 4151, with the goal of sampling different source states and fluxes.  Systematic studies including all of those observations, and studies focused on variability in emission lines and winds, will be submitted over the coming year.  A number of these observations will make use of complementary data from the Xtend CCD camera, which is even more important for spatially extended sources.

\section{Conclusions}
In the broadest terms, our result indicate that:\\

\noindent(1) The narrowest velocity components of Fe~K emission lines in unobscured AGN likely arise in neutral gas and/or dust, at negligible bulk velocity.\\

\noindent(2) The Fe~K ``line'' is actually a composition of emission from distinct parts of the accretion flow.  In particular, the line in NGC 4151 has velocity components that are concretely associated with the inner wall of the torus, the innermost extent of the optical BLR (or, ``X-ray BLR''), and potentially a geometry closer to the central engine that may signal a warped accretion disk.  This fulfills the unique promise of Fe~K$_{\alpha}$ lines to trace all of the key elements of the accretion geometry in AGN.\\

\noindent(3) The absence of strong Compton shoulders in narrowest components of the Fe~K line complex in NGC 4151 may suggest Compton-thin gas, consistent with a disk atmosphere and/or winds generated at the leading edge of geometrical boundaries rather than in optically thick regions.  Some of the iron flux may be due to the irradiation of dust grains, qualitatively consistent with models that rely on dust cross-sections to lift gas in the BLR above the plane of the disk\citep{czerny2015}.\\

\noindent(4) Monitoring of NGC 4151 with Resolve has the potential to reveal specific lags in distinct components.  If this can be extended to a small foundational sample, it may eventually be possible to estimate black hole masses via fits to single-epoch Fe~K line profiles in surveys undertaken with integral field unit calorimeters like the Athena/X-IFU \citep{barret2018}.\\


We thank the anonymous referee for an insightful and constructive review that improved this work.  We thank Jelle de Plaa for assistance with SPEX.  We acknowledge helpful conversations with Misty Bentz, Steve Kraemer, and Missagh Mehdipour.  We thank Gerard Kriss for sharing STIS data and for helpful conversations.  We thank the directors and scheduling and operations teams of Chandra, NICER, XMM-Newton, NuSTAR, and Swift for coordinated observations that will be important in subsequent papers.  Part of this work was performed under the auspices of the U.S. Department of Energy by Lawrence Livermore National Laboratory under Contract DE-AC52-07NA27344. The material is based upon work supported by NASA under award number 80GSFC21M0002. This work was supported by JSPS KAKENHI grant numbers JP22H00158, JP22H01268, JP22K03624, JP23H04899, JP21K13963, JP24K00638, JP24K17105, JP21K13958, JP21H01095, JP23K20850, JP24H00253, JP21K03615, JP24K00677, JP20K14491, JP23H00151, JP19K21884, JP20H01947, JP20KK0071, JP23K20239, JP24K00672, JP24K17104, JP24K17093, JP20K04009, JP21H04493, JP20H01946, JP23K13154, JP19K14762, JP20H05857, and JP23K03459. This work was supported by NASA grant numbers 80NSSC20K0733, 80NSSC18K0978, 80NSSC20K0883, 80NSSC20K0737, 80NSSC24K0678, 80NSSC18K1684, and 80NNSC22K1922. LC acknowledges support from NSF award 2205918. CD acknowledges support from STFC through grant ST/T000244/1. LG acknowledges financial support from Canadian Space Agency grant 18XARMSTMA. MS acknowledges the support by the RIKEN Pioneering Project Evolution of Matter in the Universe (r-EMU) and Rikkyo University Special Fund for Research (Rikkyo SFR). AT and the present research are in part supported by the Kagoshima University postdoctoral research program (KU-DREAM). SY acknowledges support by the RIKEN SPDR Program. IZ acknowledges partial support from the Alfred P. Sloan Foundation through the Sloan Research Fellowship. This work was supported by the JSPS Core-to-Core Program, JPJSCCA20220002. The material is based on work supported by the Strategic Research Center of Saitama University.

\bibliography{main}{}
\bibliographystyle{aasjournal}

\clearpage

\begin{figure}[t]
    \centering
     \includegraphics[width=1.0\columnwidth]{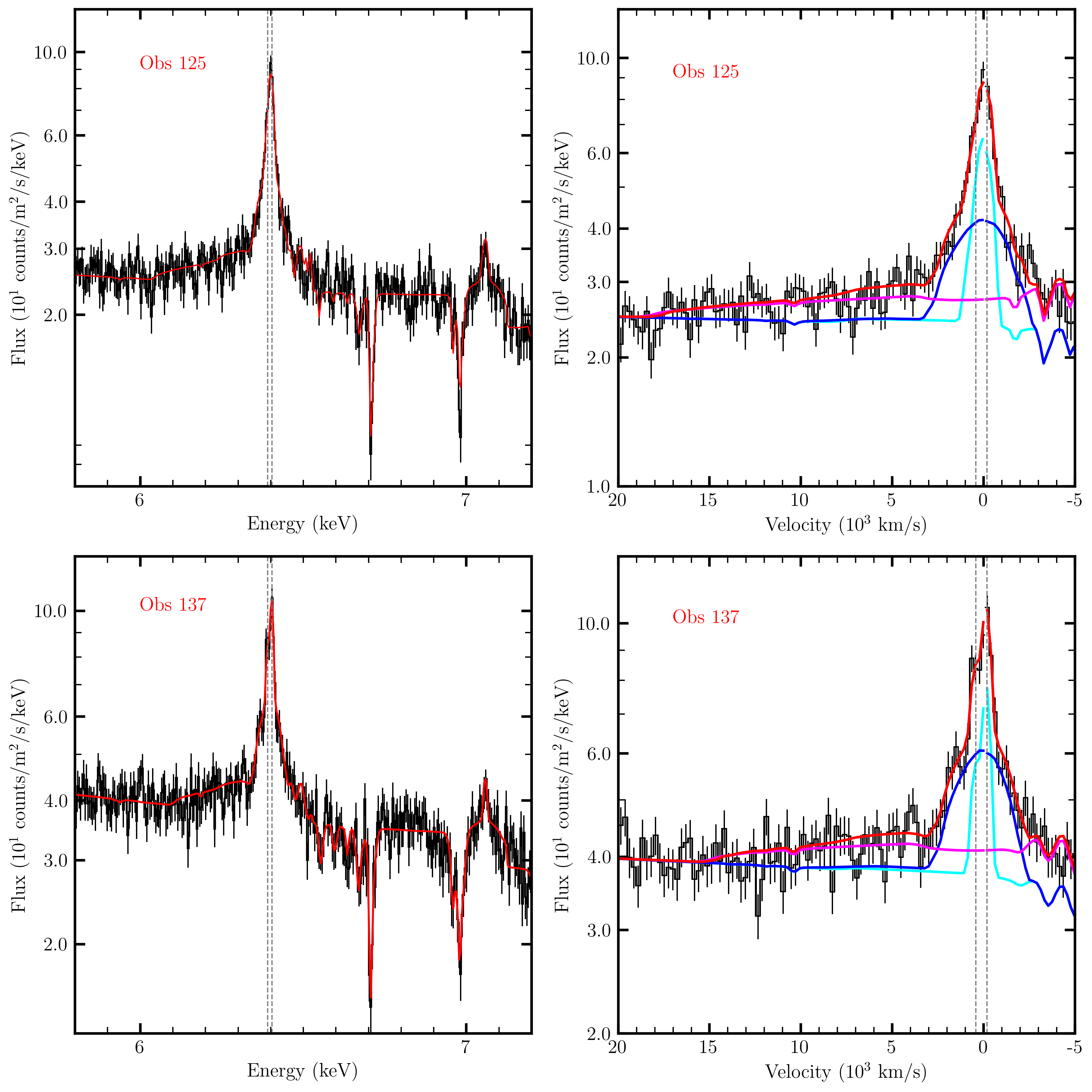}
    \caption{XRISM/Resolve spectra of NGC 4151 from Observation 000125000 and Observation 000137000.  The spectra are binned for visual clarity, and shifted to the frame of the host.  The plots in energy space cover a slightly broader range than the plots in velocity space, which aim to reveal the details of the complex Fe~K lines better.  The velocity-space plots show narrow (cyan), intermediate (blue), and broad (magenta) line components, likely from the innermost face of the torus, inner face of the BLR, and a warp or discontinuity at approximately $r\simeq 100~GM/c^{2}$, respectively.  The best-fit model for each spectrum is shown in red, including absorption lines from X-ray warm absorber winds that will be treated in a forthcoming paper (Xiang et al., in prep).  The dashed vertical lines mark the location of Fe~K$_{\alpha,1}$ and K$_{\alpha,2}$ line componenst at 6.404~keV and 6.391~keV, respectively.}
    \label{fig:fig1}
\end{figure}

\begin{figure}[t]
    \centering
     \includegraphics[width=1.0\columnwidth]{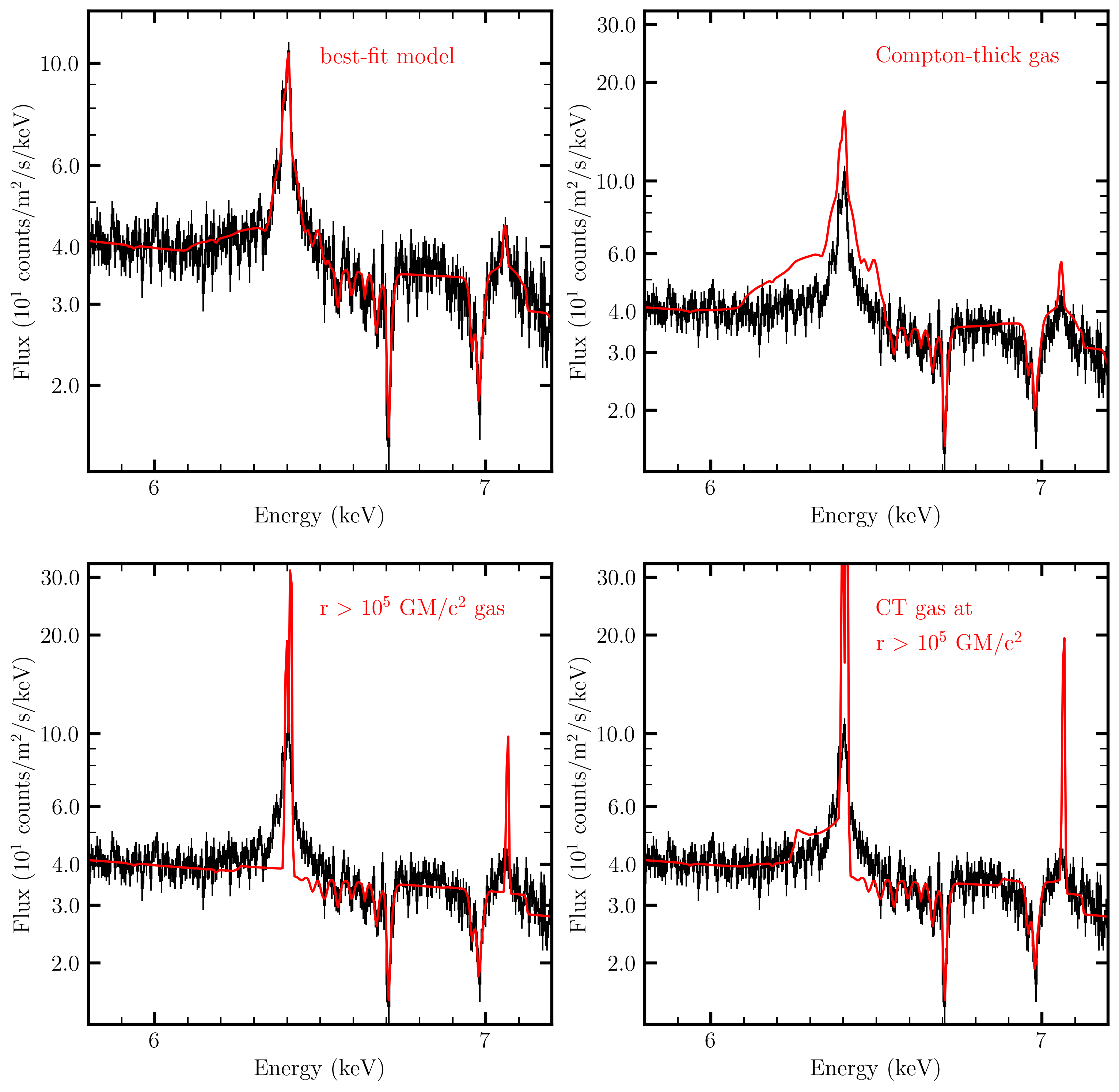}
    \caption{XRISM/Resolve spectra of NGC 4151 from Observation 000137000, illustrating the effect of changing key Fe~K line model parameters.  The data do require Compton shoulders, as determined by the models; these feature are just weak and blurred by orbital motions.  Top left: the overall best-fit model, as detailed in Table 1 and Figure 1.  Top right: the best-fit model, but altered by setting the column density of all three line components to be Compton-thick ($N_{H} = 1.6\times 10^{24}~{\rm cm}^{-2}$).  Bottom left: the best-fit model, but altered by setting the inner radius of all three components to $r = 1\times 10^{5}~GM/c^{2}$.  Bottom right: the best fit model, but with both modifications.  The line flux normalizations were unaltered in creating the comparison models.}
    \label{fig:fig2}
\end{figure}

\begin{table}[htb]
\caption{Neutral Fe~K Line Complex Model Parameters}
\begin{scriptsize}
\begin{center}
\begin{tabular}{lll}
Parameter & Obs. 000125000 & Obs. 000137000\\
 ~        & (Dec. 02 2023) & (Dec 26 2023) \\
\tableline
$r_{nar}~(GM/c^{2})$ &  $5600^{+5000}_{-3100}$  &  $21000^{+46000}_{-10000}$   \\
$\theta_{nar}~{\rm (deg.)}$ & $12.0^{+8.2}_{-4.6}$  & $10.2^{+7.5}_{-3.8}$  \\
${\rm N}_{\rm H}~(10^{24}~{\rm cm}^{-2})$ &  $0.22^{+0.90}_{-0.20}$ &  $0.17^{+0.15}_{-0.16}$  \\                
Norm. $(10^{7})$  & $1.53^{+2.7}_{-0.6}$  &  $1.40^{+0.13}_{-0.26}$\\
\tableline
$r_{intm}~(GM/c^{2})$  &   $3100^{+1400}_{-1200}$ &  $2800^{+500}_{-700}$ \\
$\theta_{intm}~{\rm (deg.)}$ &   $33.7^{+27.3}_{-7.7}$ &  $24.8^{+8.2}_{-5.1}$ \\
${\rm N}_{\rm H}~(10^{24}~{\rm cm}^{-2})$&  $0.18^{+1.20}_{-0.17}$ &  $0.17^{+0.59}_{-0.17}$  \\
Norm. $(10^{7})$ &  $2.1^{+0.2}_{-0.2}$ & $3.2^{+0.1}_{-0.2}$  \\
\tableline
$r_{brd}~(GM/c^{2})$ &  $73.8^{+103.2}_{-35.3}$ &  $97.0^{+96.3}_{-9.8}$  \\
$\theta_{brd}~{\rm (deg.)}$ &  $19.4^{+1.3}_{-0.2}$ &  $19.5^{+0.8}_{-0.5}$ \\
${\rm N}_{\rm H}~(10^{24}~{\rm cm}^{-2})$ & $0.1^{+9.9}_{-0.10}$ &  $0.09^{+0.26}_{-0.08}$  \\
Norm. $(10^{7})$ &  $3.2^{+10.0}_{-1.2}$ & $3.2^{+4.4}_{-0.2}$  \\
\tableline
$\Gamma$ &  $1.63^{+0.01}_{-0.01}$  &   $1.71^{+0.01}_{-0.02}$ \\
Norm. $(10^{7})$ &  $1.80^{+0.05}_{-0.05}$  &  $3.20^{+0.17}_{-0.11}$ \\                 
\tableline
${\rm N}_{\rm H}~(10^{24}~{\rm cm}^{-2})$  &  $0.17^{+0.05}_{-0.02}$ &  $0.14^{+0.26}_{-0.04}$ \\
$f_{cov}$ &  $0.96^{+0.04}_{-0.16}$ & $0.84^{+0.16}_{-0.4}$  \\
\tableline
$C/\nu$ & 474/400 & 406/432  \\
\tableline
\end{tabular}
\vspace*{\baselineskip}~\\
\end{center} 
\tablecomments{Key spectral model parameters values and $1\sigma$ errors.  The Fe~K complex was fit with a series of three additive ``mytorus'' line components (narrow, intermediate, and broad), modified by Keplerian broadening via the ``Speith'' model within SPEX.  The ``mytorus'' component includes K$_{\alpha,1}$ and K$_{\alpha,2}$ lines, and corresponding K$_{\beta}$ lines.  Via ``Speith,'' we measured the inner radius for the component and its inclination (assuming a standard Euclidean emissivity of $J\propto r^{-3}$), as well as the column density of the emitting gas and the flux normalization of each component.  These line models were added to an underlying cut-off power-law that was modified by partial covering absorption within the nucleus or ISM of NGC 4151 via that ``hot'' absorption model within SPEX.  The line normalizations are defined in the same units as the power-law normalization, which has units of $10^{44}~{\rm ph}~{\rm s}^{-1}~{\rm keV}^{-1}$ at 1~keV.  All fits were made in the 5.40--7.40~keV band by minimizing a Cash statistic after binning by a factor of 9.0.  Please see the text for other key details of the spectral models and procedure.} 
\end{scriptsize}
\end{table}

\begin{figure}[t]
    \centering
     \includegraphics[width=1.0\columnwidth]{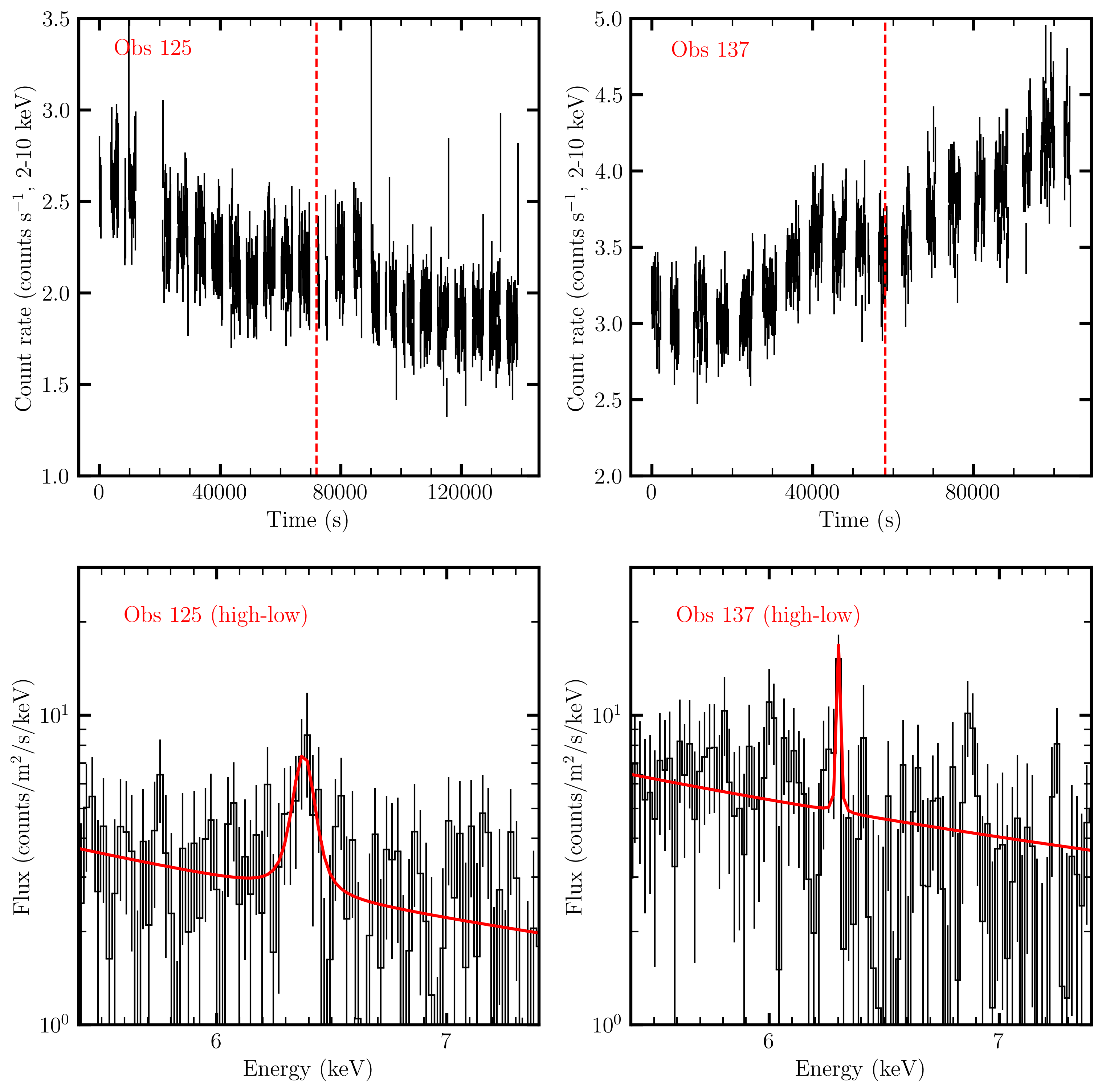}
    \caption{XRISM/Resolve light curves of NGC 4151 in the 2--10~keV band, and high--low flux ``difference'' spectra.  The difference spectra were created by dividing the observations into two equivalent time intervals representing relatively high flux and low flux phases, creating spectra from each, and subtracting the low flux phase from the high flux phase.  The resultant difference spectrum reveals components that vary on the given time scale.  The variable Fe~K$_{\alpha}$ flux in Observation 000125000 is significant at the 5$\sigma$ level.  The narrow, tentative Fe~K$_{\alpha}$ line flux in Observation 000137000 is marginally significant at the 3$\sigma$ level of confidence.  The 50--70~ks intervals sampled in these crude selections correspond to distances of $r = c\Delta t \simeq 300-400~GM/c^{2}$ assuming direct light travel paths in NGC 4151, and distances a factor of a few smaller allowing for geometrical effects.  Please see the text for additional details.}
    \label{fig:fig3}
\end{figure}

\begin{figure}[t]
    \centering
     \includegraphics[width=0.45\columnwidth]{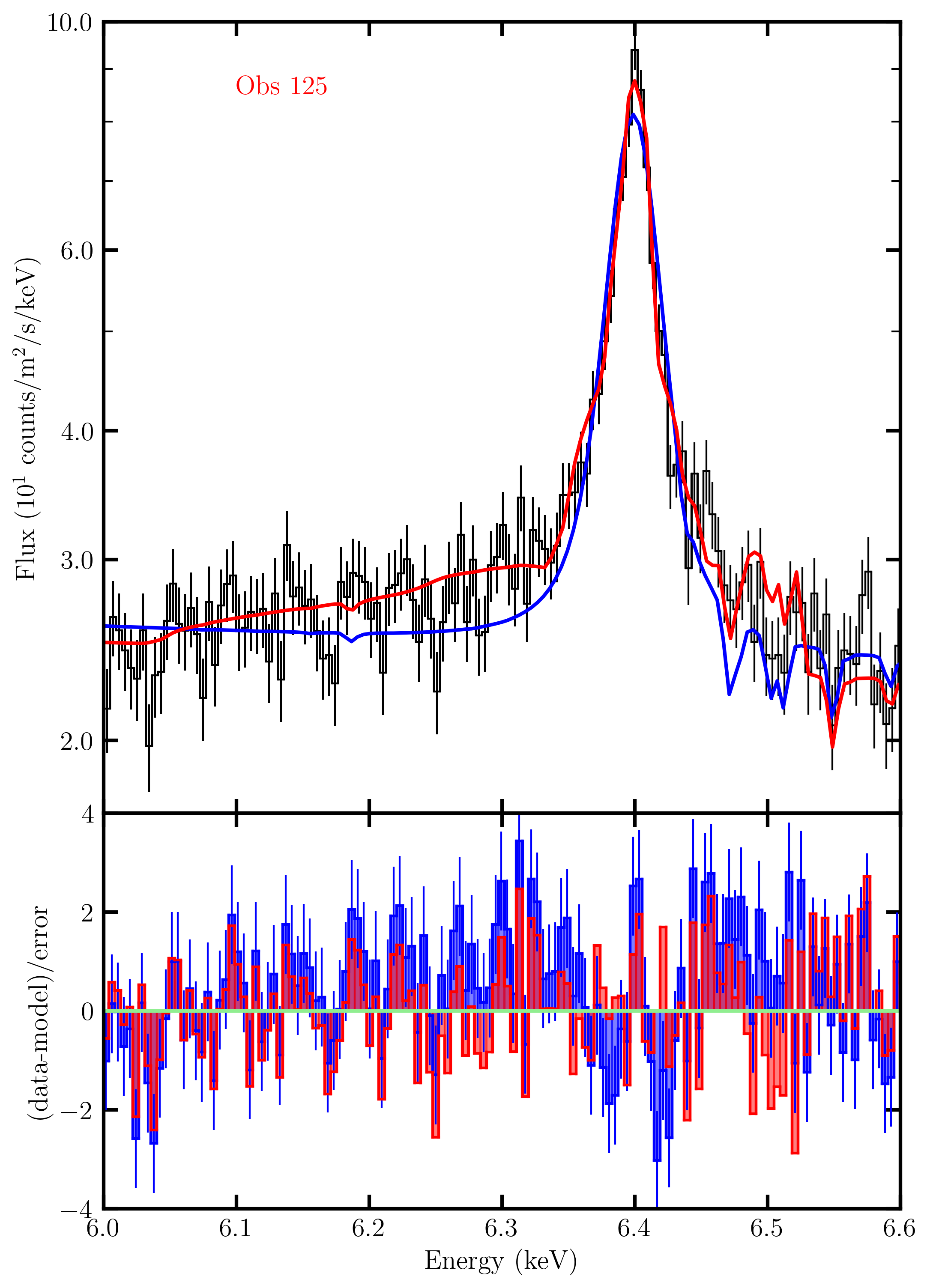}
     \includegraphics[width=0.45\columnwidth]{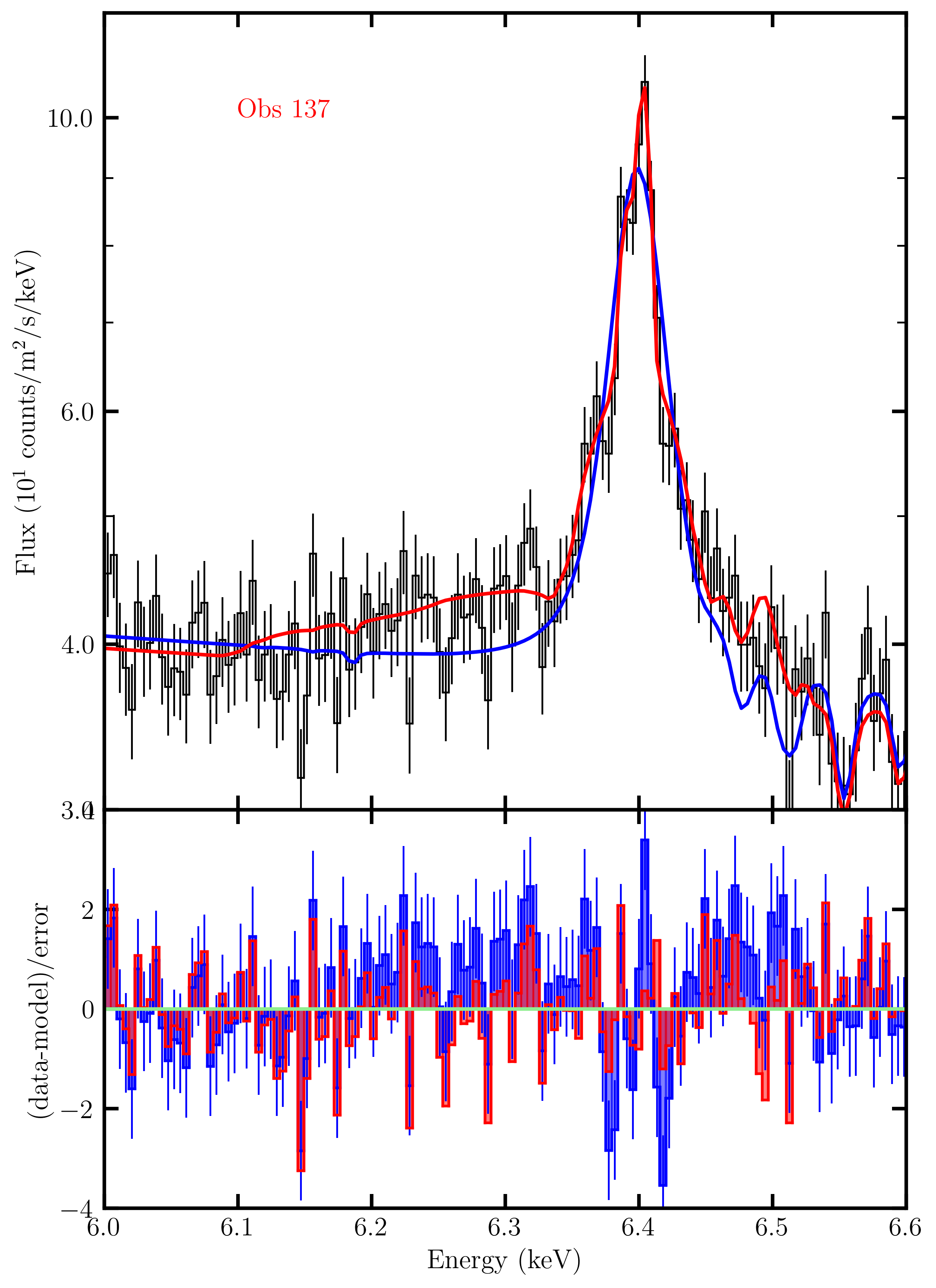}
    \caption{XRISM/Resolve spectra of NGC 4151 focusing on the narrow Fe~K line.  The best fit model for each spectrum is shown in red, and an alternative fit with a Voigt function is shown in blue.  This effectively tests the need for three components since the Voigt function is a composition of Gaussian and Lorentzian profiles.  The bottom panel of each plot traces the quality of the models in corresponding colors.  In both spectra, the combination of three components provides a significantly better fit to the narrowest component, the intermediate component, and red wing to the line profile.  Treating the Cash statistic as a $\chi^{2}$ statistic in a high-count limit, the best-fit three-component models detailed in Table 1 statistically preferred over the Voigt functions at more than the 8$\sigma$ level of confidence.  Please see the text for additional details.  For clarity, error bars are only retained on the blue data in the bottom panels; the red data points have equivalent errors.}
    \label{fig:fig4}
\end{figure}

\begin{figure}[t]
    \centering
     \includegraphics[width=1.0\columnwidth]{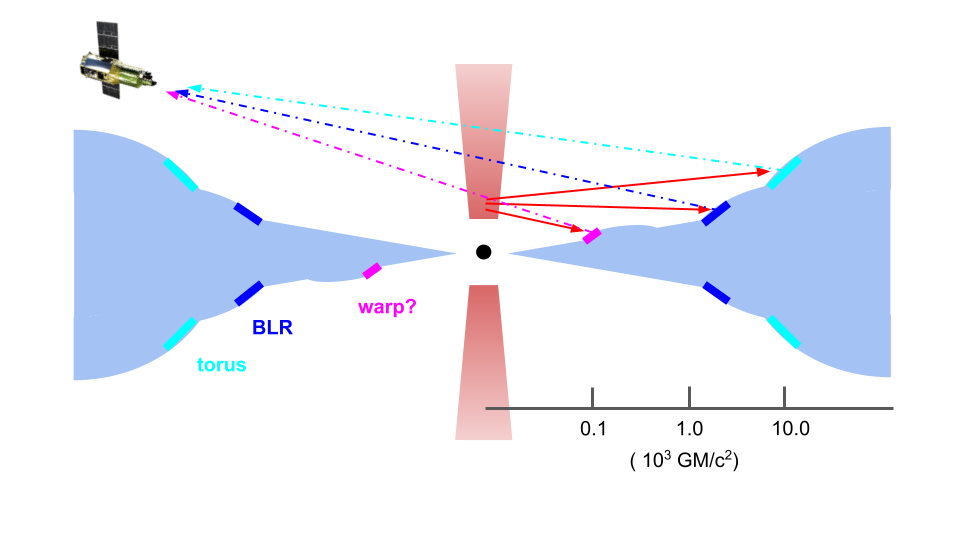}
    \caption{A simple cartoon giving a possible geometrical interpretation of the Fe~K$_{\alpha}$ line observed in NGC 4151 with XRISM.  The geometry is depicted in cross-section, with light blue indicating the disk (and disk atmosphere).  Vertical scale heights and angles are figurative, not quantitative.  The corona is depicted here as the base of an outflow, though other possibilities exist.  Hard X-rays from the corona irradiate all parts of the inflow, but the face of those regions with extra local vertical extent are irradiated more and contribute specific line flux components.  These are the magenta, blue, and cyan regions associated with a potential warp, the inner extent of the BLR, and the inner extent of the torus are indicated; these colors match the line components in Figure 1.  For simplicity and clarity, irradiation and Fe~K$_{\alpha}$ emission are only depicted from the ``far'' side of the central engine, though the full 2$\pi$ cylinder must be illuminated.  This diagram also neglects winds that are observed in absorption in NGC 4151.  Please see the text for details alternative explanations of the component associated with a warped disk in this diagram. }
    \label{fig:fig5}
\end{figure}

\end{document}